\documentclass[a4paper,11pt]{article}
\pdfoutput=1 % if your are submitting a pdflatex (i.e. if you have
\usepackage{jinstpub} % for details on the use of the package, please
\title{Calibration of the scintillation of cerium-doped yttrium aluminum garnet crystals irradiated by monoenergetic 4 MeV energy electrons}

\author[a,b,1]{K. Burdonov,\note{Corresponding author.}}
\author[a,c,1]{P. Forestier-Colleoni}
\author[d]{J-N Cayla}
\author[c]{T. Ceccotti}
\author[d]{S. Chanc\'{e}}
\author[d]{V. Chaumat}
\author[d]{N. Delerue}
\author[d]{H. Guler}
\author[a]{V. Lelasseux}
\author[d]{B. Lucas}
\author[a]{J. Fuchs,}

\affiliation[a]{LULI - CNRS, CEA, UPMC Univ Paris 06 : Sorbonne Université, Ecole Polytechnique, Institut Polytechnique de Paris - F-91128 Palaiseau cedex, France}
\affiliation[b]{IAP, Russian Academy of Sciences, 603950 Nizhny Novgorod,  Russia}
\affiliation[c]{LIDYL, Commissariat \`{a} l’Energie Atomique, DSM/DRECAM, CEA Saclay,
91191 Gif sur Yvette, France}
\affiliation[d]{IJCLab, CNRS Université Paris-Saclay, F-91405 Orsay Cedex, France}
\emailAdd{konstantin.burdonov@polytechnique.edu} \emailAdd{pierre.forestier-colleoni@polytechnique.edu}

\abstract{This paper presents the results of measurements of fluorescing cerium-doped yttrium aluminum garnet crystals after being irradiated by an accelerated electron beam with energy of around 4 MeV. The measurements were performed using the PHIL linear accelerator at LAL (France). We observe linear dependence of the crystal emission to the electron beam charge and the isotropy of the photon emission. We provide the calibration coefficients of the photon emission depending on the charge of the accelerated electron beam for two crystals originating from two different manufacturers.}


\begin{document}
\maketitle
\flushbottom

\section{Introduction}
\label{sec:intro}

Cerium-doped yttrium aluminum garnet (YAG:Ce) is a scintillation material produced in a form of single-crystal of composition Y$_{3}$Al$_{5}$O$_{12}$. It emits yellow light with maximum emission at 550 nm wavelength when subjected to ionising radiation (UV, x-rays, charged particles). Fluorescence decay time is composed of a fast decay time of 70 ns and a slower decay time of the order of $\mu$s \cite{a,Mihokova2007}. 
YAG:Ce scintillators are used in a wide range of applications, including cathode tubes, PET scanners, high-energy gamma radiation and charged particle detectors, and high-resolution imaging screens for gamma, x-rays, beta radiation and ultraviolet radiation \cite{b,c,d}.

For laser-plasma interaction experiments, such as those widely performed at petawatt-class laser facilities \cite{e}, YAG:Ce scintillators can be used as end detectors in magnetic spectrometers of ionizing radiation. With the democratisation of high repetition rate Petawatt-class laser facilities (state-of-art lasers, such as Apollon \cite{f}, Vega \cite{g} and ELI-NP \cite{h} are capable of delivering optical pulses to a target chamber with one shot per minute repetition rate), fast reusable detectors are needed to optimize the efficiency of particle and radiation detection. 

Currently single-use detectors (such as Image Plate\cite{Boutoux2016,Strehlow2019} and CR-39\cite{Jeong2017}) are used in laser-plasma interaction experiments. They are not compatible with high repetition rate as they need to be replaced after each shot. A good candidate for particles detection produced during such interactions is the YAG:Ce scintillating material as its fluorescence decay time is shorter than 1~s (90~ns). Another of its advantage is the minimum out gazing of the crystal compared to Lanex screen\cite{Buck2010,Rabhi2016} allowing their use in extremely high vacuum. Moreover its fluorescence emission is in the visible spectrum (maximal at 550~nm \cite{a,Mihokova2007}) which is suitable for regular CCD or CMOS devices coupled with an imaging system.

In this paper we present the results of YAG:Ce scintillators calibration performed at the linear accelerator of electrons (LAL, Orsay) \cite{i,j,k}. The aim here is to calculate the photon emission response from the bulk of YAG:Ce crystal exposed to an electron beam. Two individual crystals from two different manufacturers (see Table~\ref{tab:crystal_summary}) were irradiated by a collimated mono-energetic electron beam having variable energy from 3.5~MeV to 4.2~MeV and charge from 2~pC to 20~pC. The emission level was registered by BASLER CMOS cameras placed at three different angles relative to the direction of the electron beam.

The setup of the experiment will be described in section~\ref{sec:setup} followed by the primary results and a discussion in section~\ref{sec:results}. Finally, we will draw our conclusions in section~\ref{sec:sum}.

\section{Experimental setup description}
\label{sec:setup}

\begin{figure}[htp]
    \centering
    \includegraphics[width=13 cm]{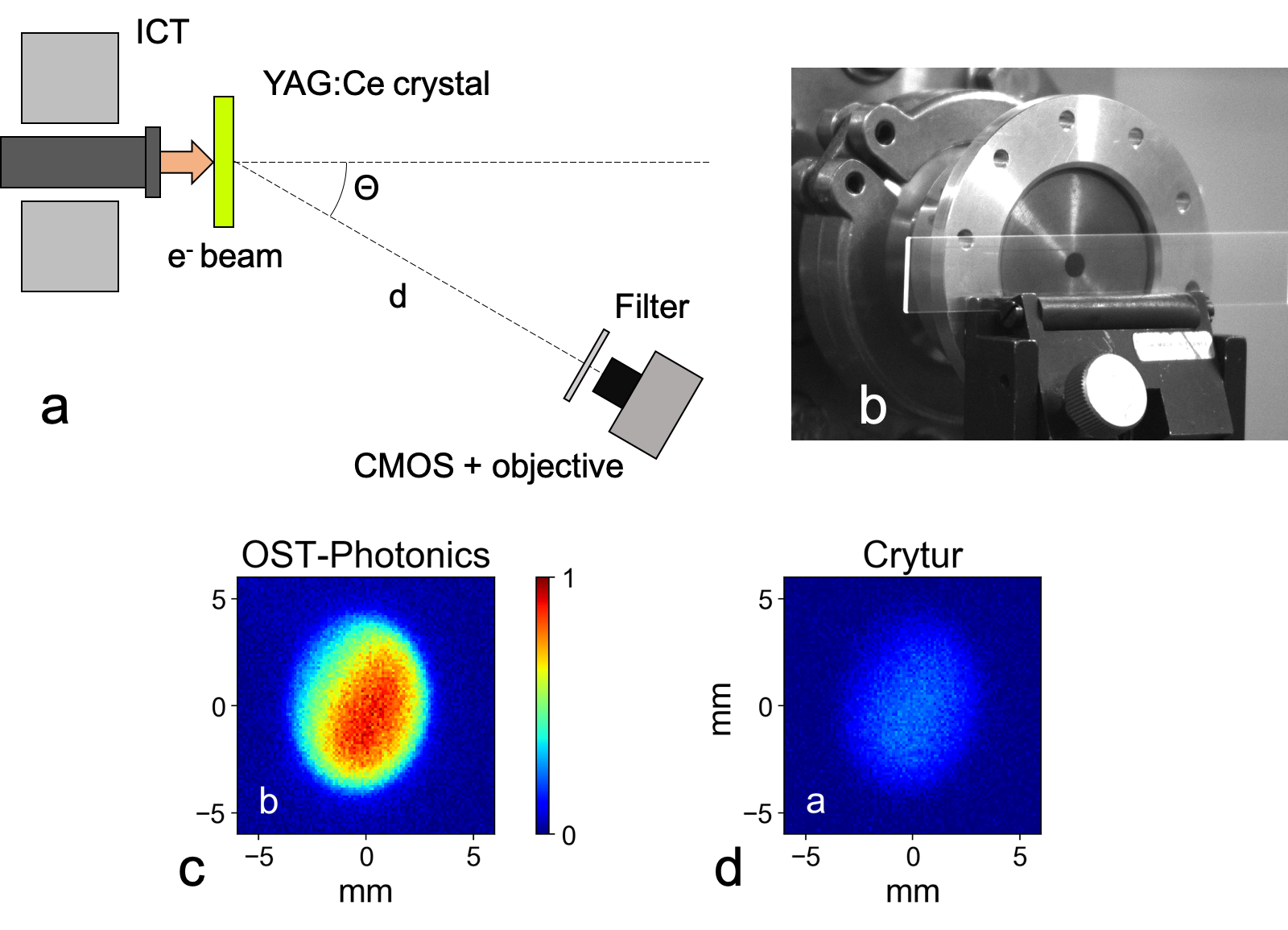}
    \caption{Schematic view of the experimental set-up (a); photo of the crystal placed in front of the titanium output window of the electron accelerator (b); normalized YAG:Ce crystal emissions registered by CMOS detector (see text)from two different manufacturer OST-Photonics (c) and Crytur (d). The signal level is normalized to the maximum brightness of the OST-Photonics one.}
    \label{fig:setup}
\end{figure}

The linear electron accelerator PHIL at LAL (Laboratoire de l'accélérateur linéaire, Orsay, France), used for the calibration, provides a collimated monoenergetic beam of accelerated electrons with 1 ps duration, charge up to 20 pC, energies from 3.45 MeV to 4.2 MeV and 5 mm diameter at the output. The setup of the experiment is shown in the Figure \ref{fig:setup}. The electron beam impacted the YAG:Ce crystal sample, which was placed on the axis of its path. Two YAG:Ce crystals from two different manufacturers (see Table~\ref{tab:crystal_summary}) were calibrated.

\begin{table}[h]
    \centering
    \begin{tabular}{c||c|c|c|c}
         & Manufacturer & Doping fraction & Thickness & Measured light yield $l_y$ \\
        \hline\hline
        Crystal 1 & OST-Photonics & 2 \% & 0.5 mm & 4800 photons/MeV \\
        Crystal 2 &  Crytur & 1 \% & 1 mm & 500 photons/MeV
    \end{tabular}
    \caption{Characteristics of the two crystals used during the experiment.}
    \label{tab:crystal_summary}
\end{table}

The light emitted from the bulk of the crystal was registered by the 8-bit Basler ACE CMOS matrix (model acA1300-30gm) equipped with an objective lens at three angles from the electron axis $\theta=10,30$ and $80^{\circ}$ and a distance $d$ of approximately $\sim$ 1 m from the crystal, corresponding to a collection solid angle of $\sim$ 25 $\mu$sr). We used a range of camera gains from minimum to maximum and two exposure times: 10~ms and 100~ms. Neutral density filters (from Thorlabs) and a narrow-band filter (central wavelength at 550 nm and 100 nm bandwidth from Edmund Optics) were used to optimize the imaging system. Examples of images obtained with this set-up are shown in Figure~\ref{fig:setup} (c, d).

The charge of the electron beam was measured by means of an Integrating Current Transformer (ICT) placed at the end of the accelerator line before the output beam collimator and the output titanium window, and by a Faraday Cup (FC) at the place of crystal mounting position (in air, after the output titanium window). The exact charge of the electron beam, impacting the YAG:Ce crystal was retrieved based on ICT measurements, with the electron transmission of the collimator and the titanium window taken into an account.

\section{Results of YAG:Ce crystals calibration}
\label{sec:results}

\begin{figure}[htp]
    \centering
    \includegraphics[width=8 cm]{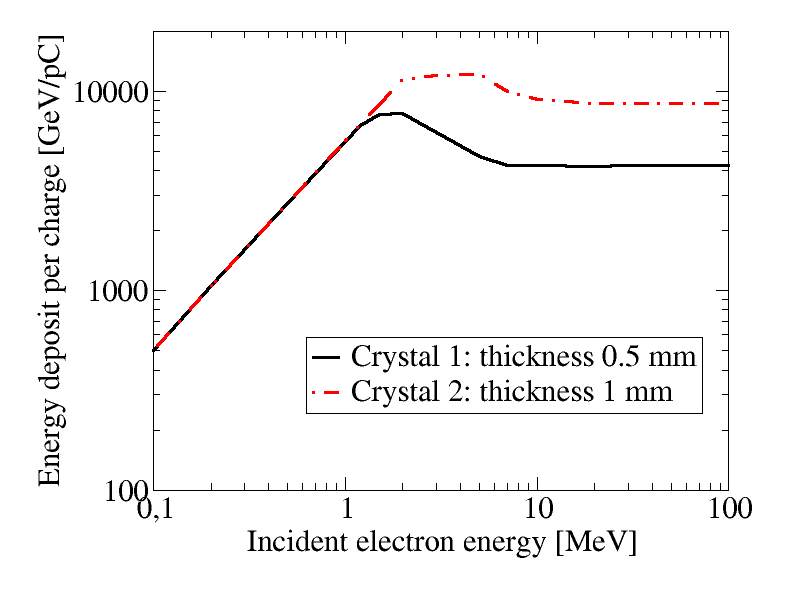}
    \caption{Geant4 calculation of the deposited energy per pC in the bulk of the YAG:Ce crystals.}
    \label{fig:geant}
\end{figure}

During the propagation of electrons inside the bulk of the YAG:Ce crystal, electrons will make collisions with atoms, losing energy. This energy will be stored in energy levels of Ce$^{3+}$.  
The excitation of the energy levels of the Ce will release photons in $4\pi$ solid angle by spontaneous emissions with two characteristics decay time 70~ns and $1~\mu$s \cite{a,Mihokova2007}.
The relation between the number of emitted photons and the incident electron charge is the crystal response to the electrons (also called light yield $L_y(\lambda)=l_y f(\lambda)$,  with $f(\lambda)$ being the normalized emission spectrum of the emitted photon that could be find in \cite{Mihokova2007,c}, $\lambda$ being the scintillation photon wavelength, and $l_y$ being the integrated light yield). It can be written as:

\begin{equation}
\label{eq:response}
N_\gamma = \int L_y(\lambda) \cdot E_{d} \cdot q ~d\lambda = l_y \cdot E_{d} \cdot q ,
\end{equation}
where $N_\gamma$ is the total number of photons emitted by the crystal, $E_{d}$ is the deposited energy of the electron beam per pC, and $q$ is the charge of the electron beam in pC. $E_{d}$ depends on the incident electron energy and can be calculated using the GEANT4 toolkit \cite{l,m,n} for different crystal compositions and thicknesses (see Figure~\ref{fig:geant}). 

The charge $q$ of the electron beam was obtained using the calibrated ICT measurement. The total number of photons emitted by the crystal $N_\gamma$ was retrieved from the integrated crystal fluorescent image $Image_{fluo}$ (Figure~\ref{fig:setup} (c, d)) using the equation~\ref{eq:nb_photon} with an assumption of isotropic photon emission: 

\begin{equation}
\label{eq:nb_photon}
Image_{fluo}-Image_{bg} = \int N_\gamma \cdot \frac{d\Omega}{4\pi}\cdot T_{f}\cdot \eta_{cmos}\,d\lambda,
\end{equation}
where $Image_{bg}$ corresponds to the signal from the crystal image without electron beam interacting with it, $d\Omega$ is the collection solid angle of the imaging system, $T_{f}$ is the transmission of filters in front of the camera, and $\eta_{cmos}$ correspond to the detector parameters, which in our case is the quantum efficiency of the Basler CMOS chip $QE_{cmos}$ multiplied by the camera Gain $\Gamma$ ($\eta_{cmos}=QE_{cmos}\cdot\Gamma$) \cite{MeasurementProt}.

From the equations~\ref{eq:response} and~\ref{eq:nb_photon}, we can retrieve the number of photons as a function of the charge:

\begin{equation}
\label{eq:CMOS_signal}
N_\gamma = \frac{4\pi}{d\Omega}\cdot\frac{Image_{fluo}-Image_{bg}}{\int T_{f}\cdot \eta_{cmos}\,d\lambda} = l_y\cdot E_d \cdot q.
\end{equation}

\begin{figure}[htp]
    \centering
    \includegraphics[width=9 cm]{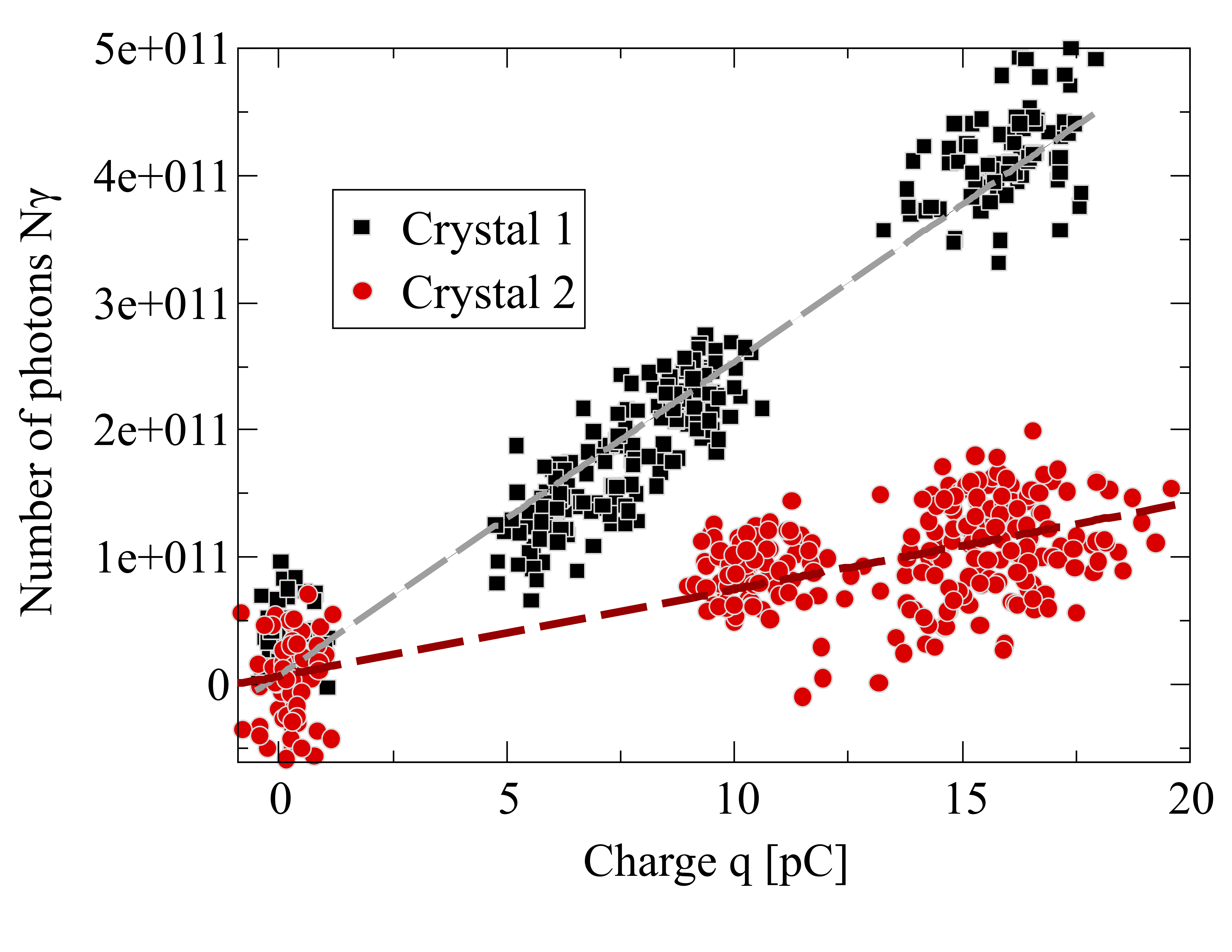}
    \caption{Integrated signal collected by the CMOS detector as a function of the total charge contained in the electron beam of 4.11 MeV. The dashed line represents the linear fit of the data for the two crystals.}
    \label{fig:CCD_vs_charge}
\end{figure}

On Figure~\ref{fig:CCD_vs_charge} we plotted the number of photons emitted by the crystal as a function of the electron beam charge for the two different crystals. It is clearly seen from the figure that the factor $a=l_y\cdot E_d$ remains constant in the range from 1 pC to 20 pC and equals to $a_1= (2.46\pm 0.2)\times 10^{10} $ photons/pC for the crystal 1 and $a_2= (6.8\pm 0.5)\times 10^{9}$ photons/pC for the crystal 2. Using Figure~\ref{fig:geant} we obtained the light yield $l_y$ for the two crystals (see Table~\ref{tab:crystal_summary} "light yield $l_y$" column).

\begin{figure}[htp]
    \centering
    \includegraphics[width=9 cm]{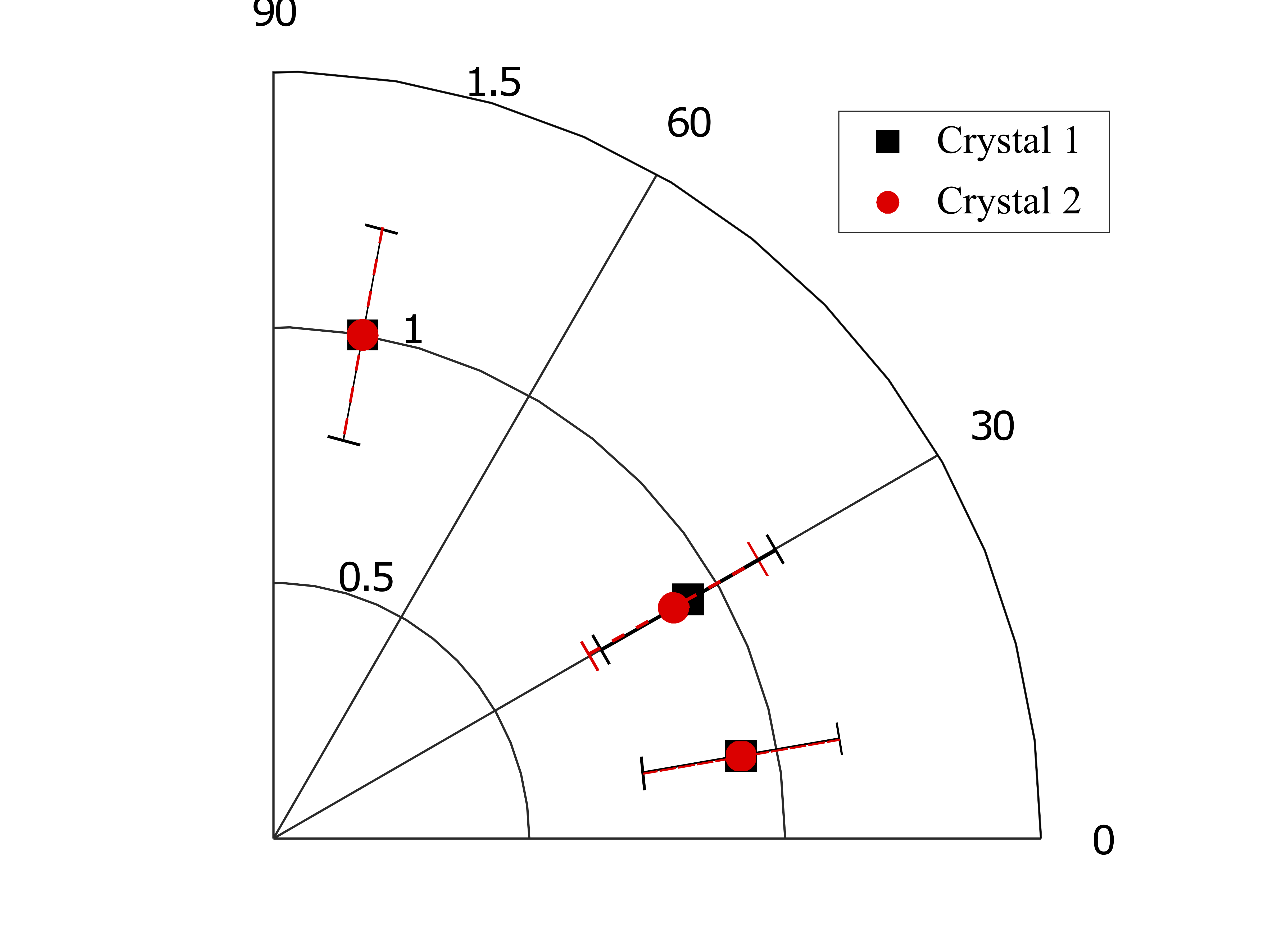}
    \caption{Polar graph of light yield depending on the CMOS detector position angle $\theta$ (see Figure \ref{fig:setup}), normalized to the CMOS response at 80$^{\circ}$. The electron beam energy was 4.11~MeV.}
    \label{fig:angular_distribution}
\end{figure}

As can be seen from Figure~\ref{fig:angular_distribution}, the signal level registered by the detector is independent of the angle $\theta$ (see Figure \ref{fig:setup}), that is in consistency with the assumption of isotropic photon emission from the crystal, used in Equation~\ref{eq:CMOS_signal}.

The measurements of the signal level with 10 times increased exposure time (100 ms) of the Basler camera did not affect the values of $a_1$ and $a_2$ factors, only increasing the grey level of the background signal (in consistency with a fluorescence decay time shorter than the acquisition time). We also changed the electron beam energy from 3.5~to 4.2~MeV. The results show similar trend and values. 
Those measurements were also performed for another Crytur's YAG:Ce crystal, showing same trend and values of its response to electrons.

\section{Summary}
\label{sec:sum}

We performed the calibration of two similar YAG:Ce crystals from two different manufacturers. We observe that the response to electron is linear, that is ideal for high repetition rate electron or positron detection. In the studied range of electron charge, we did not observe saturation of the light emitted by the crystal. 
Our main concern is the difference of light yield from these two crystals. 
The difference between these two factors $l_y$ is close to a factor of 10. This difference cannot be directly related to the doping fraction of cesium inside of the crystal as it only a factor of 2 (see Table~\ref{tab:crystal_summary}). As these crystals are from two different manufacturers, we can think of different growth processes. 
This highlights the necessity of performing light yield $l_y$ calibration for each crystal manufacturer as the response of them can be very different.

\acknowledgments

This work was supported by funding from the European Research Council (ERC) under the European Unions Horizon 2020 research and innovation program (Grant Agreement No.787539) and by funding from the Russian Science Foundation (RSF) in the frame of project No.20-62-46050.

\end{document}